\documentclass[a4paper,11pt]{article}
\usepackage{siunitx}
\usepackage{geometry}
\usepackage[version=4]{mhchem}
\usepackage{graphicx}
\usepackage{caption}
\usepackage{subcaption}
\usepackage{lineno}
\usepackage[dvipsnames]{xcolor}
\usepackage{authblk}

\author[1]{Tecla Gabbrielli}
\author[1,*]{Natalia Bruno}
\author[2]{Nicola Corrias}
\author[1,3]{Simone Borri}
\author[1]{Luigi Consolino}
\author[4]{Mathieu Bertrand}
\author[4]{Mehran Shahmohammadi}
\author[4]{Martin Franckié}
\author[4]{Mattias Beck}
\author[4]{Jérôme Faist}
\author[1,2]{Alessandro Zavatta}
\author[1,$^\dagger$]{Francesco Cappelli}
\author[1,3,$^\dagger$]{Paolo De Natale}

\affil[1]{CNR-INO -- Istituto Nazionale di Ottica, Largo Enrico Fermi 6, 50125 Firenze FI, Italy  \newline \& LENS -- European Laboratory for Non-Linear Spectroscopy, Via Nello Carrara 1, 50019 Sesto Fiorentino FI, Italy}
\affil[2]{QTI s.r.l., Largo E. Fermi 6, 50125 Firenze, Italy}
\affil[3]{Istituto Nazionale di Fisica Nucleare (INFN), Sezione di Firenze, 50019 Sesto Fiorentino, Florence, Italy}
\affil[4]{Institute for Quantum Electronics, ETH Zürich, Zürich, Switzerland}

\affil[*]{natalia.bruno@ino.cnr.it}
\affil[$^\dagger$]{P. De Natale and F. Cappelli jointly supervised this work. }
\begin{document}


\title{Intensity correlations in quantum cascade laser harmonic frequency combs}

\maketitle


\begin{abstract}
A novel study on harmonic frequency combs emitted by Quantum Cascade Lasers (QCLs) is here presented, demonstrating the presence of intensity correlations between twin modes characterising the emission spectra. These originate from a Four-Wave Mixing (FWM) process driven by the active medium's third-order non-linearity. The study of such correlations is essential for the engineering of a new generation of semiconductor devices with the potential of becoming integrated emitters of light with quantum properties, such as squeezing and entanglement. Starting from experimental results, the limits of state-of-the-art technology are discussed as well as the possible methodologies that could lead to the detection of non-classical phenomena, or alternatively improve the design of QCLs, in the compelling perspective of generating quantum correlations in mid-infrared light.
\end{abstract}

\section{Introduction}
In the worldwide demand for modeling and developing quantum technologies, the study of novel laser sources capable of emitting non-classical light plays a crucial role for its applications in quantum information processing~\cite{lu2021advances}. Indeed, during the last decades, the intensity noise of the light emitted by semiconductor lasers has been studied in depth, also in the perspective of controlling it at the quantum level~\cite{li1998generation,yamamoto1995photon,Gensty:05}. A fundamental step for these studies is understanding the mechanism of operation of different laser devices, focusing on the different contributions that affect the intensity noise of the emitted light. Among semiconductor devices, Quantum Cascade Lasers (QCLs) deserve a particular mention, presenting an interesting possibility for developing novel quantum technologies operating in a spectral region, namely the mid-to -far infrared, so far much less explored than the telecom wavelength range. QCLs are unipolar chip-scale semiconductor lasers~\cite{Faist:1994}, nowadays well-established as sources for spectroscopy~\cite{Vitiello:2015,Consolino:2018,Borri:2019a} and actively studied for use in free-space communication~\cite{Pang:2020,Corrias:2022}. In QCLs, the active medium is a complex heterostructure composed of several semiconductor layers where the lasing transition occurs between two sub-levels of the conduction band in a cascaded multilevel system and the upper state lifetime is of the order of hundreds of {fs}~\cite{Faist:2013bo}. This internal mechanism also determines the noise contributions and the intensity noise behavior of the emitted light~\cite{Gensty:05,Tombez:2013a,schilt2015experimental, zhao2019relative}. These devices show a $1/f$-trend in the low-frequency intensity noise due to electron tunnelling through the multi-barriers structure~\cite{Borri:2011,Yamanishi:2014}. On the other hand, due to the very fast upper-state lifetime, QCLs' high-frequency noise is typical of class A lasers~\cite{khanin2012principles}. Due to their complex conduction band structure, observing quantum features in the light emitted by QCLs is difficult even when the driving current is squeezed~\cite{rana2002current}. However,
in recent years, a new motivation for the investigation of quantum properties of these sources has come from the discovery of the mechanism that is responsible for the frequency-comb generation in QCLs (QCL-combs)~\cite{hugi:2012,Friedli:2013,Riedi:2015,Faist:2016,Opacak:2019}, namely Four-Wave Mixing (FWM). This is due to the large third-order nonlinearity characterizing the active medium of these lasers~\cite{Faist:2016}. In the context of multimodal emission due to spatial hole burning, FWM guarantees a well-defined phase relation among the modes~\cite{Cappelli:2016,CappelliConsolino:2019,Consolino:2020}. Interestingly, FWM is also able to induce non-classical intensity correlations among the different modes, already observed in other materials and devices such as optical fibers and microresonators~\cite{Yuen:79,levenson:1985,slusher:1985squeezing,mccormick:2007,Dutt:2015}, but never studied in QCLs, so far. The study of this phenomenon motivates this work, giving a perspective on the possibility of observing such correlations in Mid-Infrared (MIR) QCL-combs. In particular, we demonstrate classical intensity correlations in harmonic combs emitted by MIR QCLs, characterized by a few lasing modes separated by dozens of times the free-spectral range~\cite{kazakov2017self,wang2020}, which is advantageous for spatial mode separation and analysis.Compared to standard QCL-comb emission characterized by hundreds of modes, despite its simplicity, we consider the harmonic regime suitable for the investigation, indeed having the minimum number of modes for an FWM experiment removes the ambiguities on the origin of the correlations. It thus makes the experiment much easier to be interpreted and useful for future research. Therefore, we explore harmonic QCL-combs as a test bench for studying intensity correlations. In particular, we measure them in a state characterized by a primary emission mode (the pump) and two weaker sidebands~\cite{PhysRevA.94.063807}. Ideally, if this emission is entirely due to the FWM non-linear process, the two sidebands should be equally distant from the central mode (energy conservation) and they should exhibit, in absence of laser gain or losses, intensity squeezing. In a real system, such as a QCL, additional noise sources must also be taken into account~\cite{Gensty:05}, besides losses and linear gain~\cite{Caves82}. These noise contributions can hide partially, or even completely, the correlations. In this framework, the detection of intensity correlations at the classical level is therefore the first experimental evidence of the presence of intensity correlations ascribable to FWM between different modes emitted by a QCL-comb. It also represents a springboard for the development of theoretical models, for the fabrication of new-generation  devices favoring quantum correlations, as well as suitable instrumentation for such studies. In this work, we analyze the limits and perspectives of the available devices and detection system, with the goal of evaluating future experimental implementations capable of unveiling quantum correlations.

\section{Measurement technique}%
\label{sec:experimentalSetup}

\begin{figure}[!htbp]
\centering
\includegraphics[width=0.8\columnwidth]{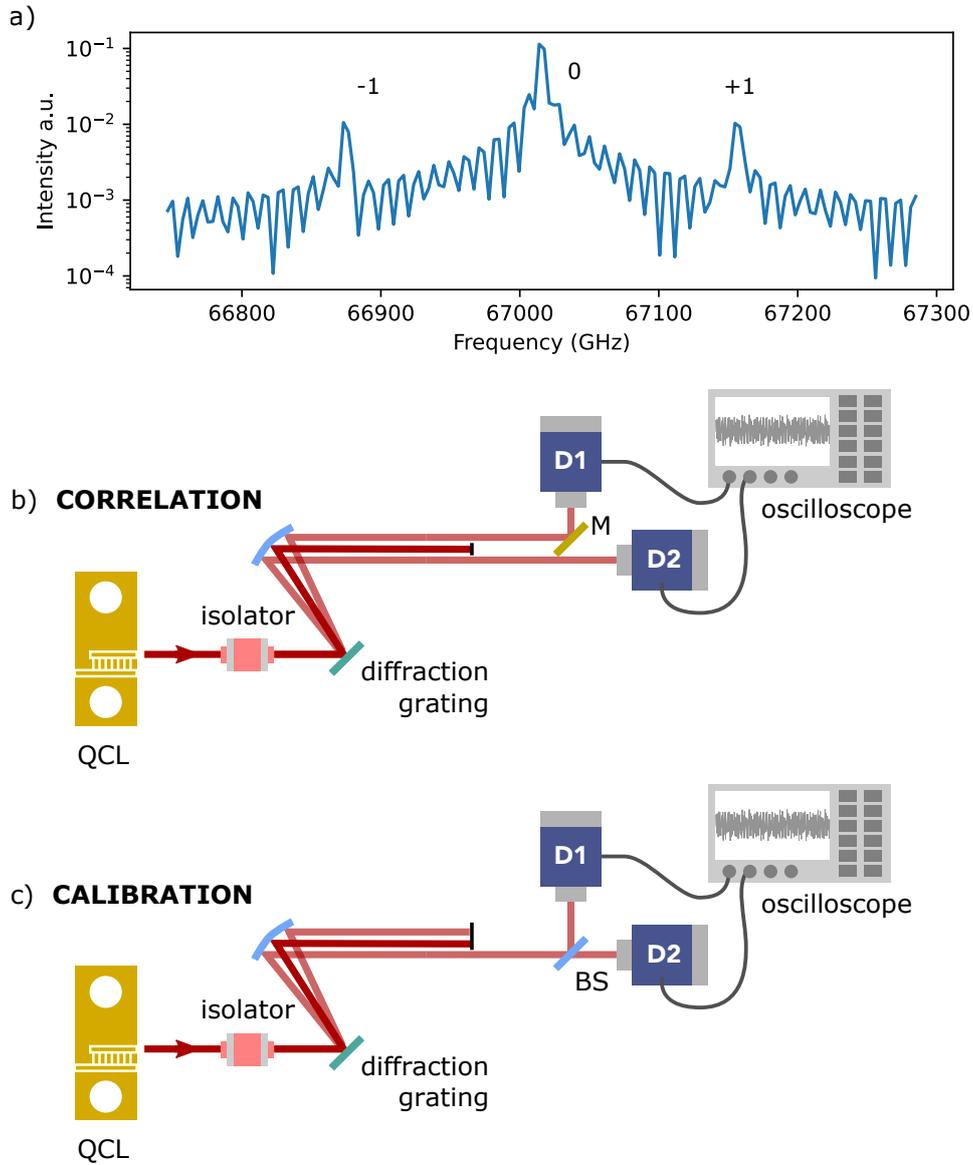}
\caption{(a) An example of the three-mode emission by the Fabry-Pérot QCL-comb emitting at \SI{4.5}{\micro\meter}. The spectrum is measured via an optical spectrum analyzer. The fluctuations observed in the detected intensity are due to the limited measurement resolution  {of \SI{6}{GHz}}. (b) Sketch of the experimental setup for correlation measurements. The three-mode emission from the QCL-comb is split with a grating. The signal of the two selected modes is sent to different photovoltaic detectors, while the third one is stopped. The signals of the two modes are acquired simultaneously in the time domain via two different channels of an oscilloscope. Data analysis is performed digitally in post production. (c) Sketch of the experimental setup for noise calibration, where we replaced the mirror $M$, shown in (b), with a 50/50 beam-splitter (BS).  {The isolator, installed in both the setups (b,c), provides isolation of \SI{40}{dB} at the laser wavelength.  With this level of isolation, the laser emission is stable and no significant feedback effect is observed.}\label{fig:setup}
}
\end{figure}

In this work, we tested QCLs emitting three-mode harmonic frequency combs, where the central, intense, mode is expected to act as pump in a FWM process generating two weaker sidebands, as shown in \textbf{Figure~\ref{fig:setup}(a)}. With the experimental setup depicted in \textbf{Figure~\ref{fig:setup}(b)} we aimed at characterizing the intensity correlations between the modes. The three modes are spatially separated by means of a diffraction grating with a groove density of \SI{300}{grooves/mm}, allowing the detection of two selected modes (e.g. the two sidebands, or one sideband and the central mode), each on one of two preamplified photovoltaic detectors (PVI-4TE-5-2x2-TO8-wAl203-36+MIP-10-250M by Vigo System). The detectors have already been extensively characterized in terms of responsivity, quantum efficiency, saturation level, and background noise in a previous work~\cite{Gabbrielli:2021}. Their main characteristics are a quantum efficiency up to 41\%, and a maximum clearance (ratio between the observable shot-noise level and the dark-noise level) of 8, leading to an effective quantum efficiency of 36\%~\cite{Gabbrielli:2021}. The detectors' DC output provides the mean value of the incident radiation intensity and the corresponding shot-noise level, to be compared to the intensity noise measured at the AC outputs. In this experiment, the two photovoltaic detectors operate in the linear response regime and the output current is therefore directly proportional to the incident power. The DC and AC output signals are acquired simultaneously in the time domain via two channels of an oscilloscope with a sample rate of \SI{625}{MS/s}. The duration of each acquisition is \SI{1}{ms}. The overall bandwidth of the setup is \SI{120}{MHz} and the Common Mode Rejection Ratio (CMRR) is up to \SI{30}{dB}~\cite{Gabbrielli:2021}.

Three different devices developed by ETH Zurich and emitting at a wavelength around \SI{4.5}{\micro \meter} were tested with the presented setup. In particular, the data presented in this manuscript, refer to a Fabri-Pérot QCL operating at a wavelength of \SI{4.5}{\micro m}  {with a waveguide length of \SI{4.5}{mm}. The laser is kept at \SI{23}{\celsius}, where the lasing threshold is \SI{470}{mA}, and driven in the current range between \SI{512}{mA} and \SI{517}{mA}.}   {In this condition, the laser operates in the harmonic comb regime where} the distance between two neighboring modes is of the order of \SI{100}{GHz} (Figure~\ref{fig:setup}(a)) {, about ten times the waveguide free spectral range}. By adjusting the position of the mirror $M$ with respect to the detector (Figure~\ref{fig:setup}(b)), it is possible to match the free spectral range of the laser under study. In Figure~\ref{fig:setup}(a) and in the following sections, the modes are labelled with the numbers +1, -1 and 0 to refer to the two sidebands and the central mode, respectively.

\subsubsection{Single-mode noise and relative shot-noise level calibration.} 

For the noise calibration, one single mode is selected with the grating, and the  mirror $M$ in Figure~\ref{fig:setup}(b) is replaced with a 50/50 beam splitter (\textbf{Figure~\ref{fig:setup}(c)}) to equally split the single-mode intensity onto the two detectors~\cite{Gabbrielli:2021}. Thanks to the  digital data analysis described in ref.~\cite{Gabbrielli:2021}, it is possible to measure both the intensity noise and the relative shot-noise level of the detected single mode. This is achieved by computing the Intensity Noise Power Spectral Density (INPSD) of the sum and the difference of the two detectors output signals, acquired simultaneously in the time domain. It is possible to estimate the intensity incident onto the balanced detector by measuring the mean values of the DC currents, which allows for computing the expected theoretical shot-noise level~\cite{Gabbrielli:2021}. For a given grating-selected mode, we verified that the shot-noise level measured with the differential balanced detection matches the expected theoretical value. Therefore, in the correlation measurements, the reference shot-noise level for a given incident power can be directly retrieved using the mean value of the detectors' DC outputs. With the presented setup it is possible to measure the INPSD of the whole signal by removing the grating and acquiring the signal with the balanced detector. In this way, the signal is not spatially dispersed and the frequency modes are overlapped on the same spatial mode. The comparison of the intensity noise level of each single mode with that of the whole three-mode signal provides information on the relation among the modes and on the presence of correlations. In particular, if the intensity noise level of the overall three-mode emission is below that of each single-mode, it can indicate the presence of correlation between the modes~\cite{note:sim_acq}. %

\subsubsection{Intensity-correlation measurement}
With this setup in Figure~\ref{fig:setup}(b), it is possible to measure both the sideband-sideband and the sideband-pump correlations by using the grating to select the mode incident on \textit{D2} and by changing the position of the mirror $M$ to select the desired mode reflected on \textit{D1}. By selecting two modes we can detect the intensity of each of them separately on two detectors simultaneously in the time domain. By comparing the INPSD of the sum of the detectors' AC-output signals (sum INPSD) with the INPSD of the difference (difference INPSD), correlations can be detected and analyzed. 
In particular, if the level of the sum INPSD is above (below) the level of the difference INPSD, the two beams are (anti) correlated. In an ideal system, two fully correlated beams have equal AC signals, and therefore a zero difference INPSD. In practice, however, the light originating from a laser active region experiences several phenomena (such as 1/f noise, gain, gain competition, waveguide loss) in the optical path and in the detection, that may affect the intensity correlation and create intensity unbalancing between different modes. In addition, the background noise of the used detectors is added on top of thes INPSD contributions. Thus demonstrating quantum-level correlations requires that the difference INPSD signal is sub-shot-noise level (below the reference shot-noise level) and that the detection system has a clearance above one~\cite{Gabbrielli:2021,Dutt:2015}. When the shot-noise level is equal to or below the background noise of the detectors, there are no chances to unveil non-classicality.
%
%
\begin{figure}[!htbp]
\includegraphics[width=0.48\columnwidth]{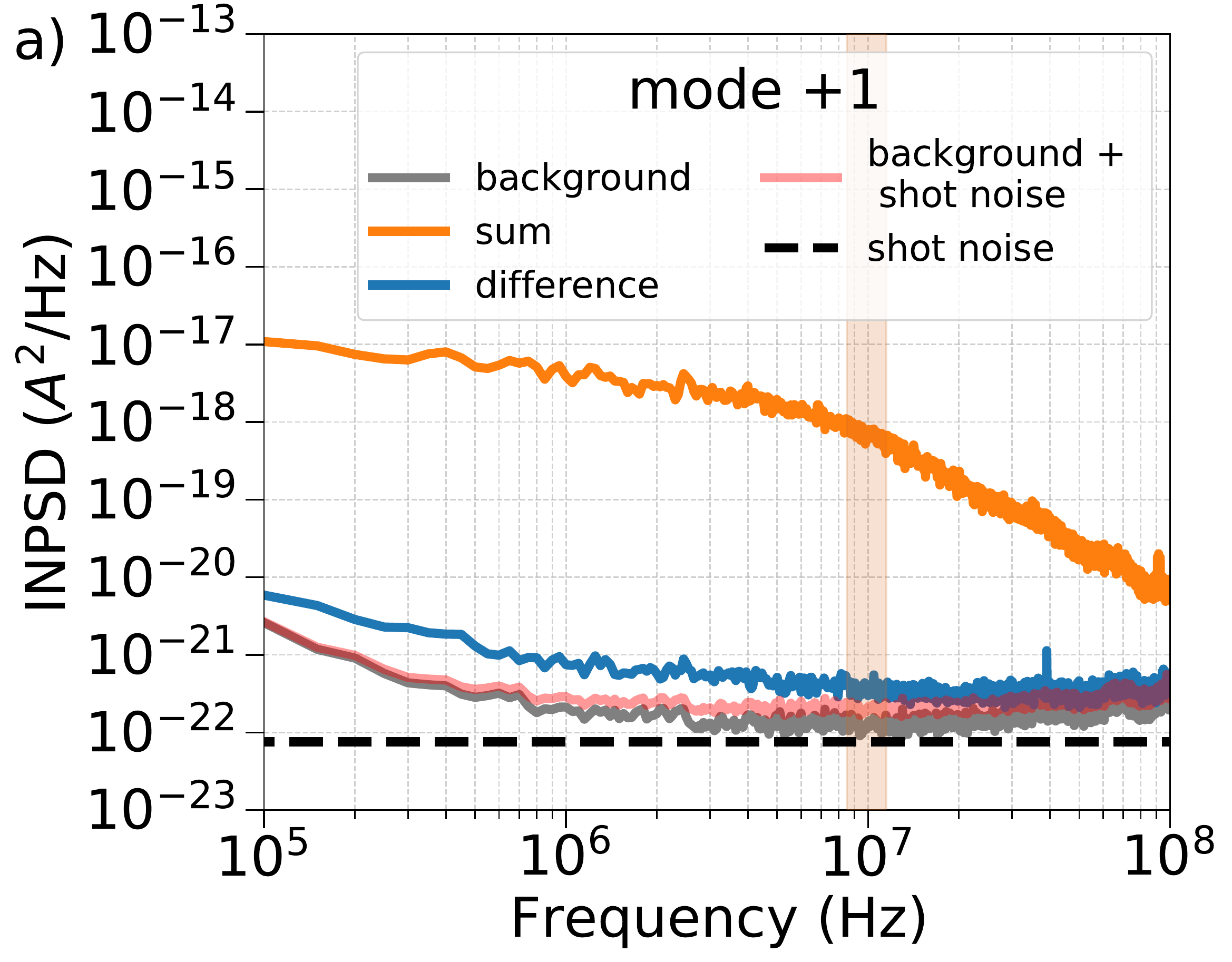} \quad 
\includegraphics[width=0.48\columnwidth]{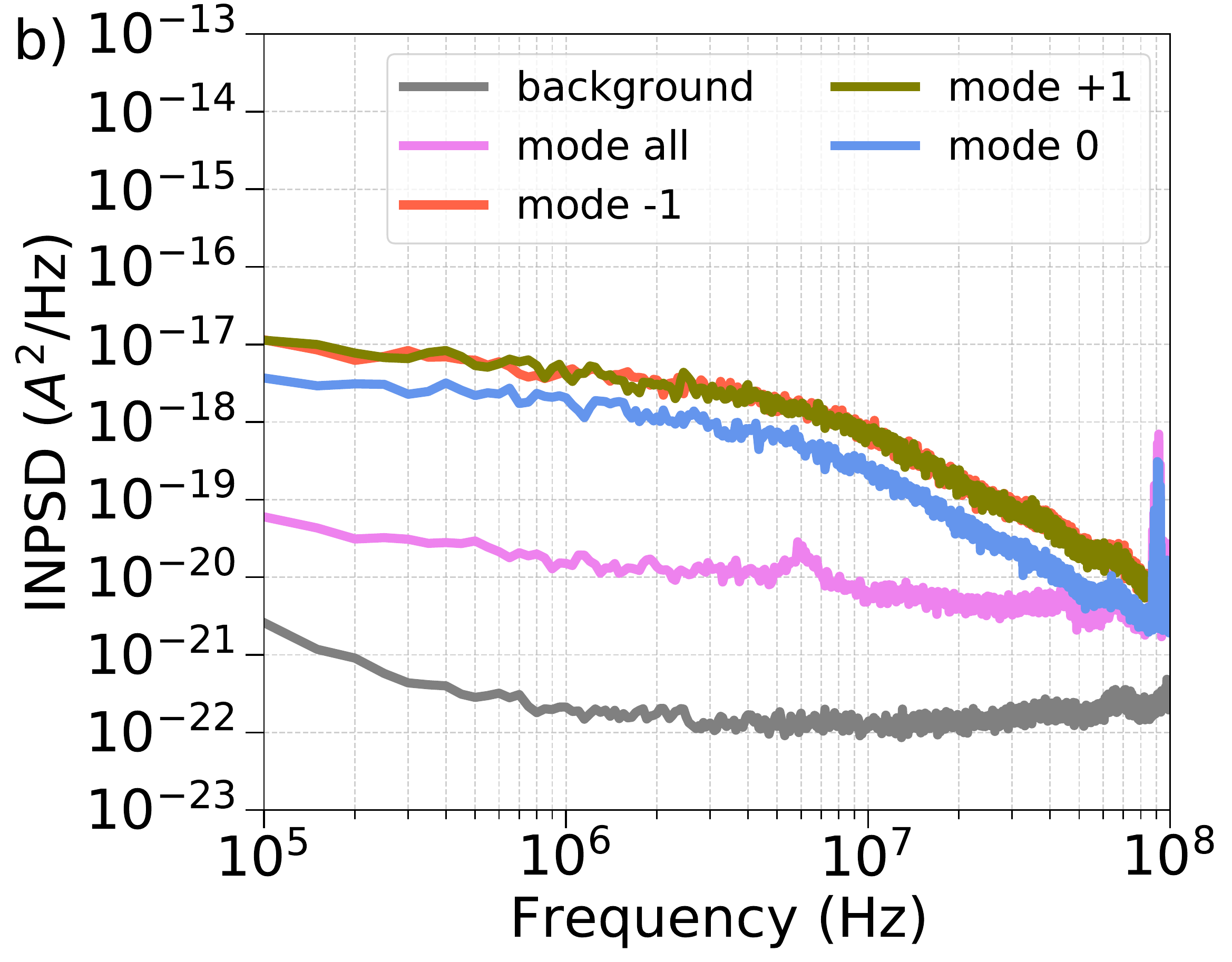}
\caption{a) INPSD of the single sideband "mode +1" detected via the 50/50 balanced detector. The INPSD of the sum (orange trace) is compared to the measured INPSD of the difference (blue trace), to the computed level of the shot noise (black dashed lines) and to the background noise of the detector. The achievement of the shot-noise level is limited by the background noise. The sum of the expected shot-noise level and of the background noise (light red trace) is compatible to INPSD of the difference. b) INPSD of each individual mode of the three-mode emission Fabry compared to the INPSD of the whole signal sent to the balanced detector without modal dispersion (no grating). +1 and -1 denote the two sidebands, while 0 the central mode. Around \SI{100}{MHz} we notice the presence of spurious noise (mode all, mode 0) compatible with FM radio broadcasting signals. The data shown in these graphs refer to the Fabry-Pérot laser emitting at \SI{4.5}{\micro m}.
}
\label{fig:RIN&singlemode}
\end{figure}
\section{Results and discussion}
\subsubsection{Single-mode intensity noise and relative shot-noise level calibration}
As described in \textbf{Section~\ref{sec:experimentalSetup}} and in Figure~\ref{fig:setup}(c), to measure the intensity noise and the corresponding shot-noise level of a single mode, we computed the INPSD of the sum and of the difference of the two detectors' output signals, respectively,  (orange and blue traces, \textbf{Figure~\ref{fig:RIN&singlemode}~(a)}. We remark that in the tested harmonic state, the sidebands are very weak compared to the central mode: their power typically ranges from a few \SI{}{\micro W} to hundreds of \SI{}{\micro W}, while the central mode power is of the order of dozens of \SI{}{m W}. As a consequence, during the measurement the central mode was attenuated in order not to exceed the saturation level of each single detector ($P>\SI{1}{mW}$), while we never had to attenuate the sidebands.
\begin{figure}[!htbp]
\includegraphics[width=0.48\columnwidth]{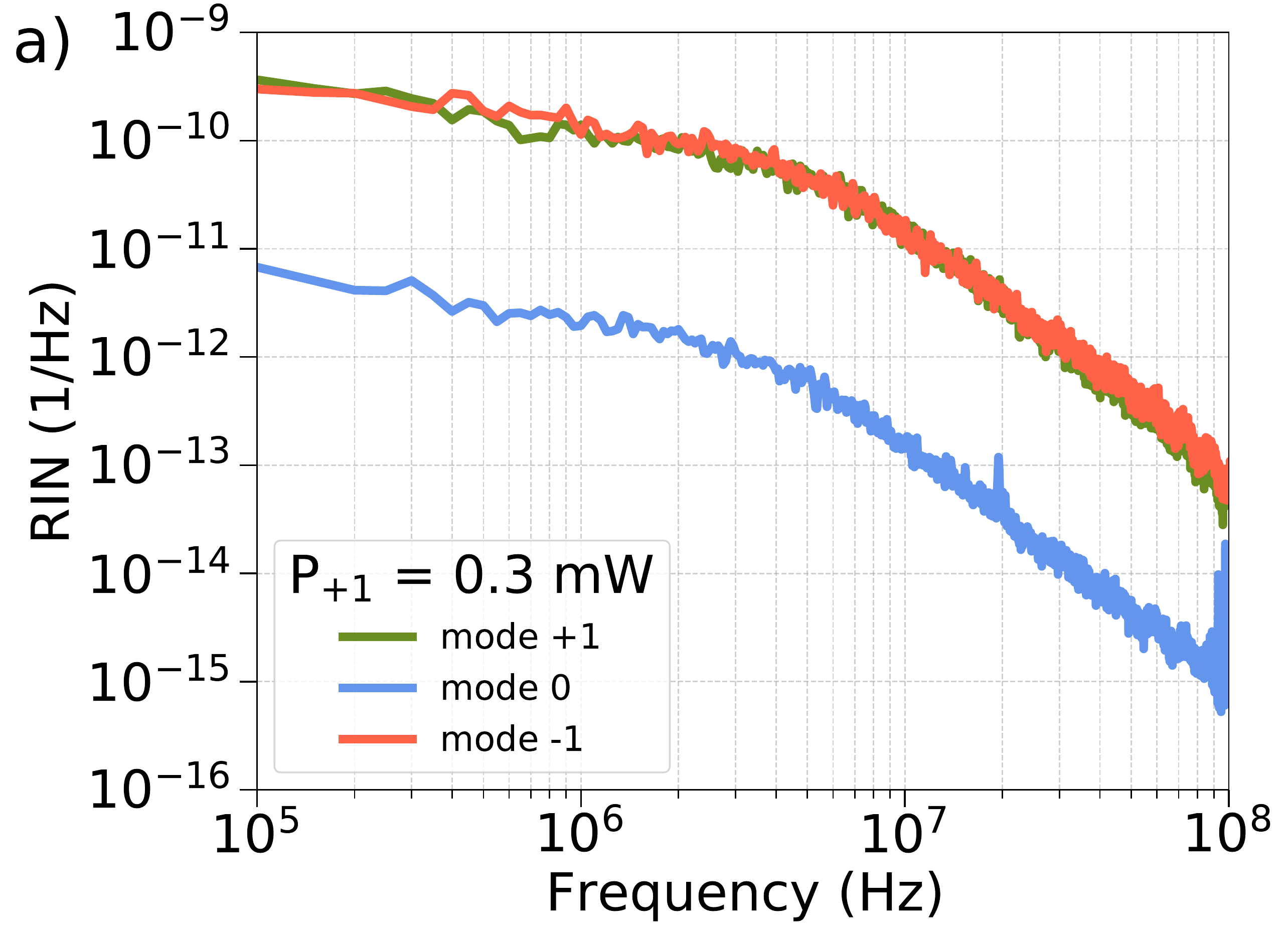} \quad 
\includegraphics[width=0.48\columnwidth]{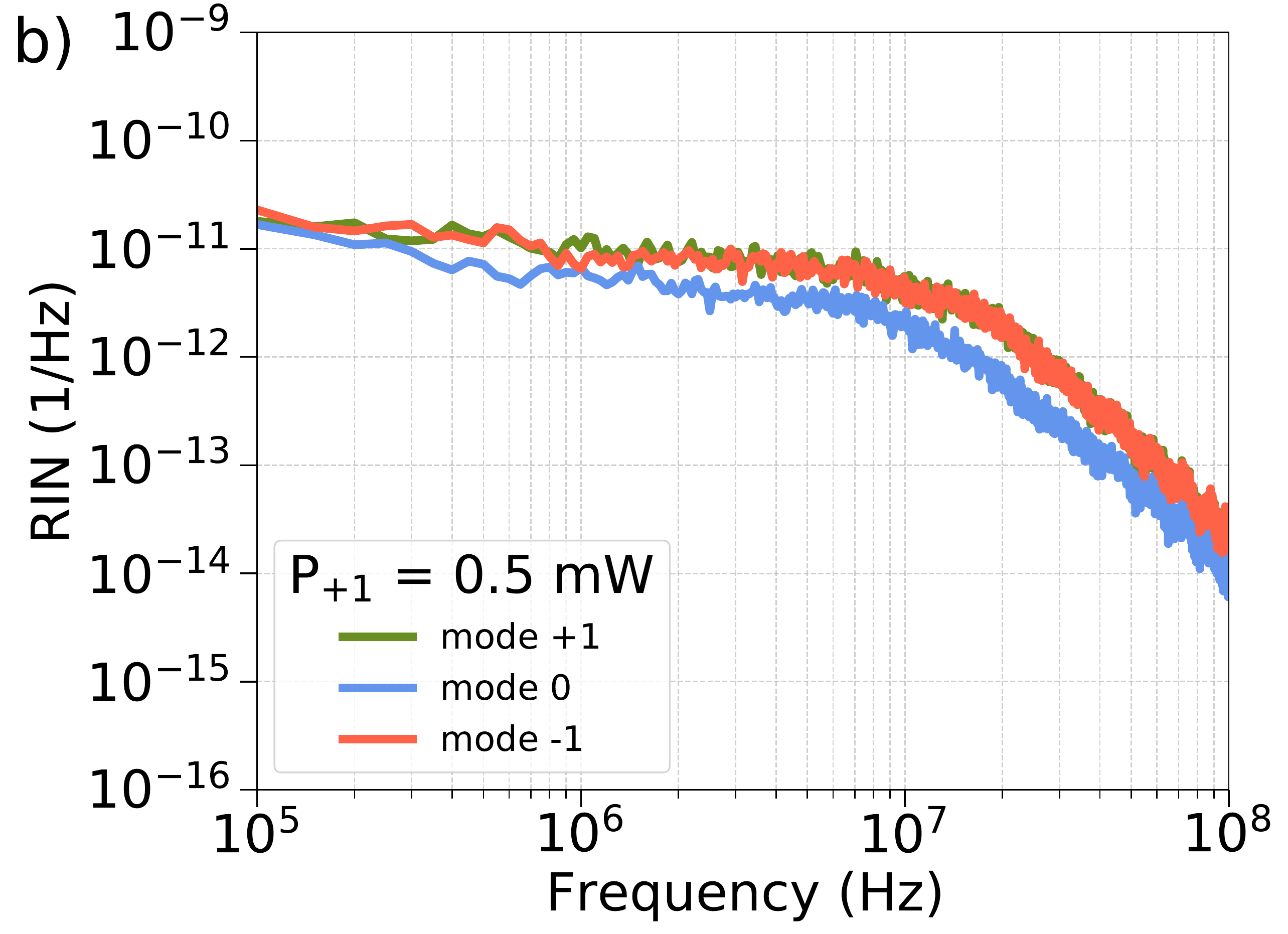}
\caption{RIN of each individual mode of the three-mode emission for two different value of the single sideband power ($\mathrm{P_{+1}}$). The balanced detector is shown in Figure~\ref{fig:setup}(c). The graph refers to the \SI{4.5}{\micro m}-wavelength QCL. The RIN of the two sidebands (mode +1, green trace and mode -1, dark orange trace) are overlapped and above the level of noise of the central mode (mode 0, blue trace). The excess noise around \SI{100}{MHz} visible in graph b) is compatible with FM radio broadcasting signals. The data shown in these graphs refer to the Fabry-Pérot laser emitting at \SI{4.5}{\micro m}.}
\label{fig:RIN}
\end{figure} 
On the other hand, the weak intensity of the sidebands translates into a shot-noise level (black dashed line) lying below the background noise (grey trace), as shown in Figure~\ref{fig:RIN&singlemode}. This limits an experimental estimation of quantum correlation, as already stated in Section~\ref{sec:experimentalSetup}. Regarding the calibration, by summing the expected shot-noise level and the background noise (light red trace in Figure~\ref{fig:RIN&singlemode}~(a)) we can properly estimate the experimental shot-noise level (blue trace in Figure~\ref{fig:RIN&singlemode}~(a)). Interestingly, this analysis evidenced that the intensity noise of the sideband (orange trace in Figure~\ref{fig:RIN&singlemode}~(a)) is well above the corresponding shot-noise level, as expected from a FWM process~\cite{mccormick:2007}.
In Figure \ref{fig:RIN&singlemode} (b), the INPSD of each sideband (modes +1 and -1, green and dark orange traces) is compared with the central mode INPSD (mode 0, blue trace) and with the INPSD of the total signal (all, pink trace). The INPSD of the total signal is lower than that of the signal of each sideband, evidencing the presence of correlation. Assuming the presence of correlation between each pair of modes, from the error theory analysis the variance $\sigma^2_{\mathrm{all}}$ of the total intensity is given by \cite{taylor1997introduction}: 
\begin{equation}
 \sigma^2_{\text{ all}} = \sigma^2_{\text{ +1}} + \sigma^2_{\text{ -1}}  + \sigma^2_{\text{ 0}} + 2 \sigma_{\text{ +1,  -1}} + 2 \sigma_{\text{ +1,0}} + 2 \sigma_{\text{0,-1}}
 \label{eq:covariance}
\end{equation}
where $\sigma^2_{\mathrm{i}}$ is the $i$-th mode variance, and $\sigma_{\mathrm{ i},\mathrm{ j}}$ is the covariance between mode $i$ and mode $j$ with i,j={+1,0,-1}. Via the Parseval theorem \cite{oppenheim1999discrete}, we can extract information about the variance of the signal by measuring its INPSD. Therefore, measuring an INPSD of the total signal (pink trace in Figure \ref{fig:RIN&singlemode} (b)) that is below each single-mode INPSD is a proof of non-negligible correlation, indeed in \textbf{Equation~\ref{eq:covariance}} the following inequality must be satisfied:
\begin{equation}
 ( \sigma_{\text{ +1,  -1}} +  \sigma_{\text{ +1,  0}} +  \sigma_{\text{ 0,  -1}} )< 0   
\end{equation} 
in order to reach a  total signal INPSD value lower than the others. \\ 
Furthermore, the analysis of the noise of each individual mode in different gain conditions (e.g. far or close to the sidebands' threshold), can give an indication of what is the most advantageous regime for the correlation measurements, as the gain is one of the parameters which can contribute to deteriorate quantum correlations, as stated in the previous sections. By changing the laser current and so the gain condition, the Relative Intensity Noise (RIN) of each single sideband (mode +1 and mode -1, \textbf{Figure~\ref{fig:RIN}}) can be higher or comparable to the RIN of the central mode (mode 0, Figure~\ref{fig:RIN}). We observed that when the sidebands are close to their threshold, they are noisier compared to the central mode (Figure~\ref{fig:RIN}(a)), while when they start to grow in power as a consequence of laser gain, their RIN decreases to the level of the central mode (Figure~\ref{fig:RIN}(b)). Interestingly, the RIN of the two sidebands is overlapped in both the configurations, evidencing the same amount of noise. The low-gain regime seems to be the most relevant condition for studying the sideband-sideband correlations, as shown in the following section. 

\subsubsection{Intensity-correlation measurement}

\begin{figure}[!htbp]
\includegraphics[width=0.48\columnwidth]{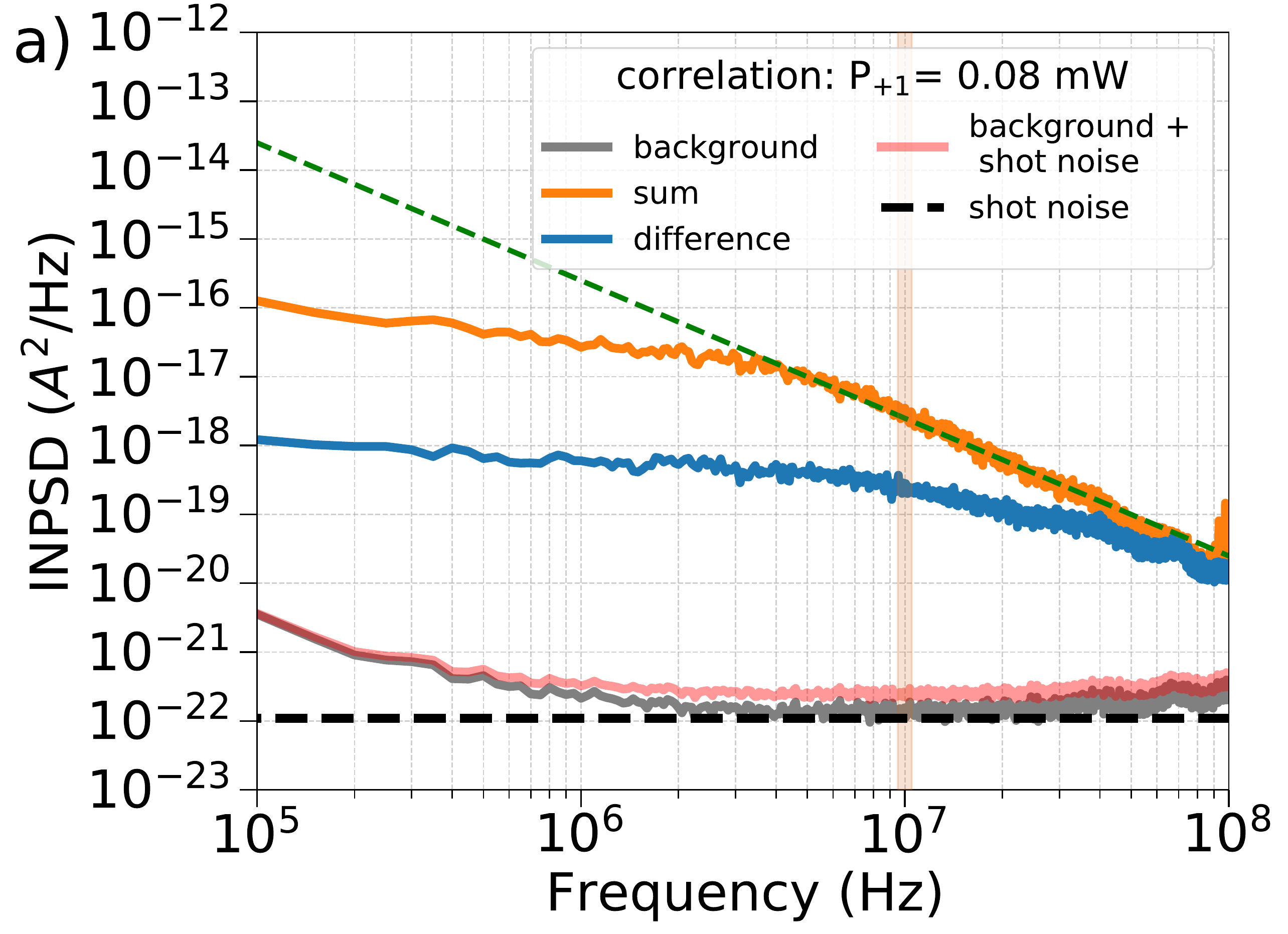}  \quad 
\includegraphics[width=0.48\columnwidth]{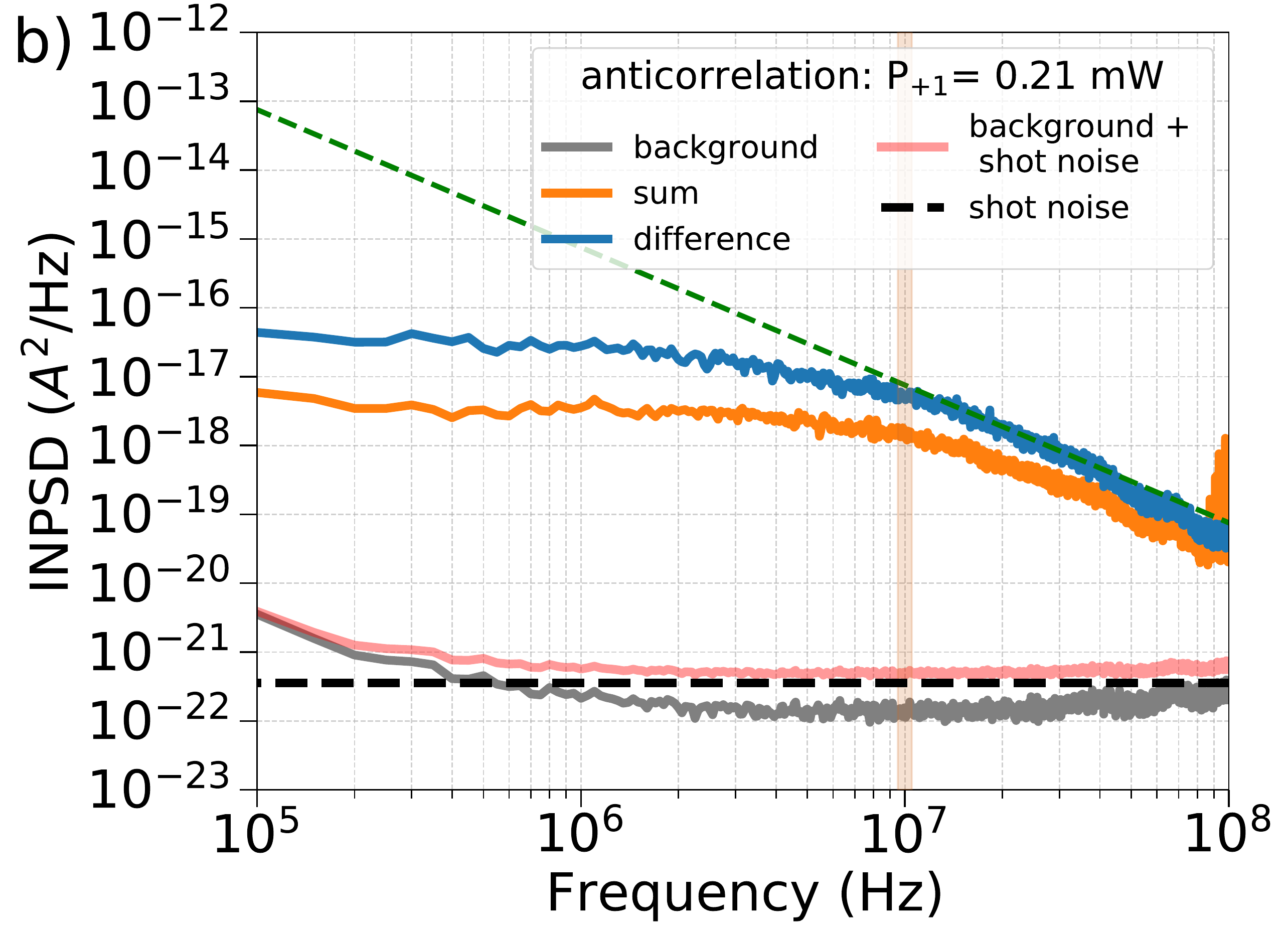}
\caption{INPSD analysis of the sideband-sideband correlation (a) and sideband-pump anti-correlation (b) for different values of the single sideband power ($\mathrm{P_{+1}}$). The INPSD of the sum (orange trace) is compared to the measured INPSD of the difference (blue trace), to the expected level of shot noise (black dashed lines) and the background noise of the detector. 
In both graphs, the brown area centered at \SI{10}{MHz} represents the Fast Fourier Transform (FFT) frequency window of \SI{1}{MHz} in which the CMRR analysis shown in \textbf{Figure~\ref{fig:CMRR}} has been performed.  { In this window, the CMRR of the setup is optimal (over \SI{30}{dB}) and the highest clearance is reached \cite{Gabbrielli:2021}.} The green dashed line indicates a frequency cut-off of the INSPD of the sum (a) and the difference (b). The data shown in these graphs refer to the Fabry-Pérot laser emitting at \SI{4.5}{\micro m}.
\label{fig:INPSD_corr_anticorr} }
\end{figure}

Using the setup shown in Figure~\ref{fig:setup}(b) and following the measurement techniques explained in Section~\ref{sec:experimentalSetup}, we investigated the sideband-sideband correlations and sideband-pump correlations. In \textbf{Figure~\ref{fig:INPSD_corr_anticorr}} an example of both trends is given at two different values of the single sideband power ($P_{+1}$). The graph (a) evidences the presence of a partial sideband-sideband correlation, with a difference INPSD that lays up to \SI{20}{dB} below the sum INPSD, but still over \SI{20}{dB} above the shot-noise level. In case of higher-intensity sidebands, a partial anti-correlation between the single sideband and the central mode is measured, as shown in Figure~\ref{fig:INPSD_corr_anticorr} (b), where the sum INPSD lays below the one of the difference. In both graphs, a green dashed line indicates the presence of a frequency cut-off in the sum INPSD and in the difference INPSD in case of correlation and anti-correlation, respectively. Interestingly, this \SI{10}{MHz} cut-off occurs clearly before the detector 3-dB-cut-off at \SI{120}{MHz}. Its presence in both the graphs is therefore not due to the detection system, and it could be related to the internal dynamics of QCL-combs, in particular to the intermodal dynamics. Indeed the cut-off is present both in the correlation spectra (see Figure~\ref{fig:INPSD_corr_anticorr}) and in the single-mode spectra (Figure~\ref{fig:RIN&singlemode} (b)), but not in the one of the whole radiation (\textit{mode all} in Figure~\ref{fig:RIN&singlemode} (b)).
\begin{figure}[!htbp]
\includegraphics[width=0.47\columnwidth]{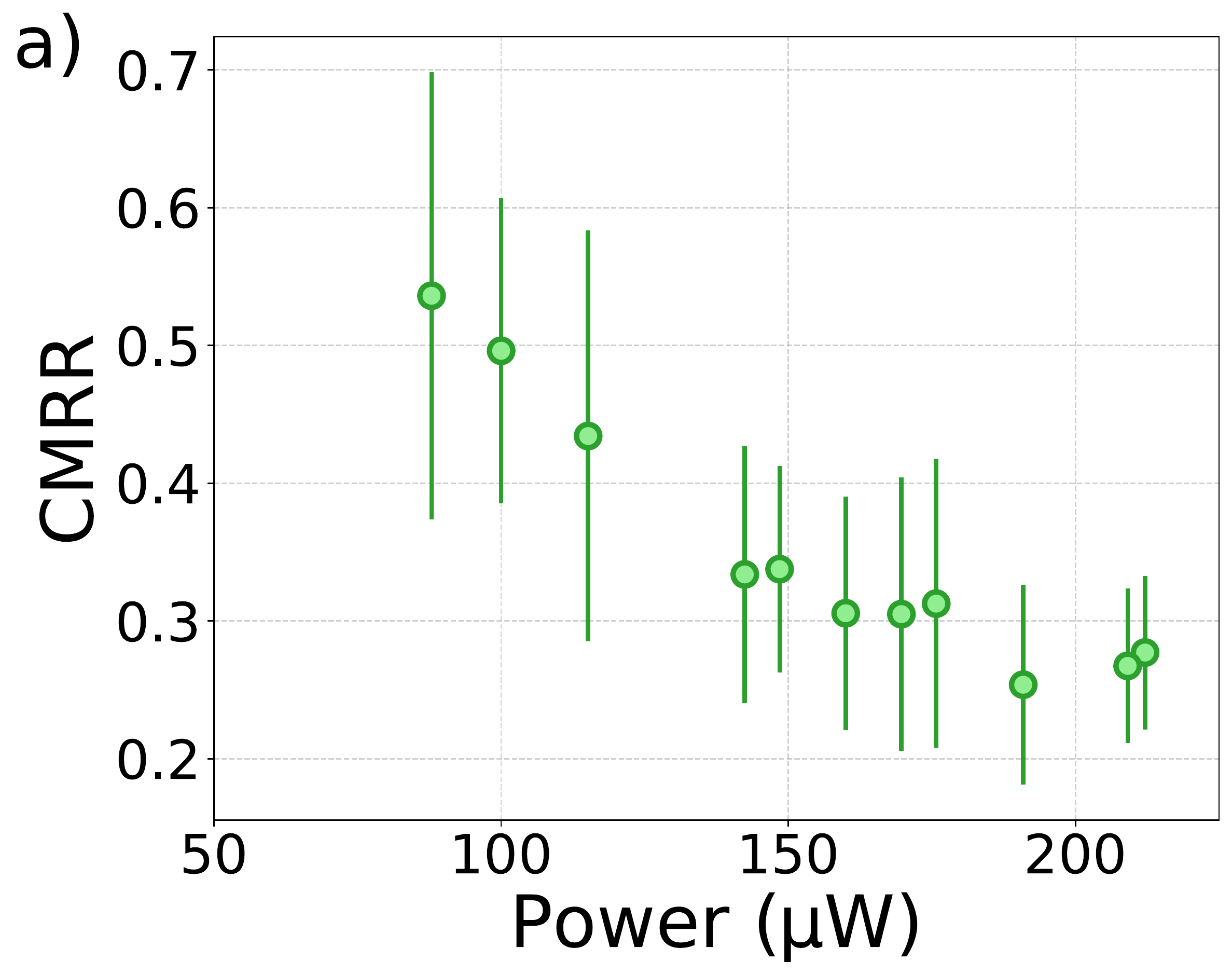} \quad
\includegraphics[width=0.48\columnwidth]{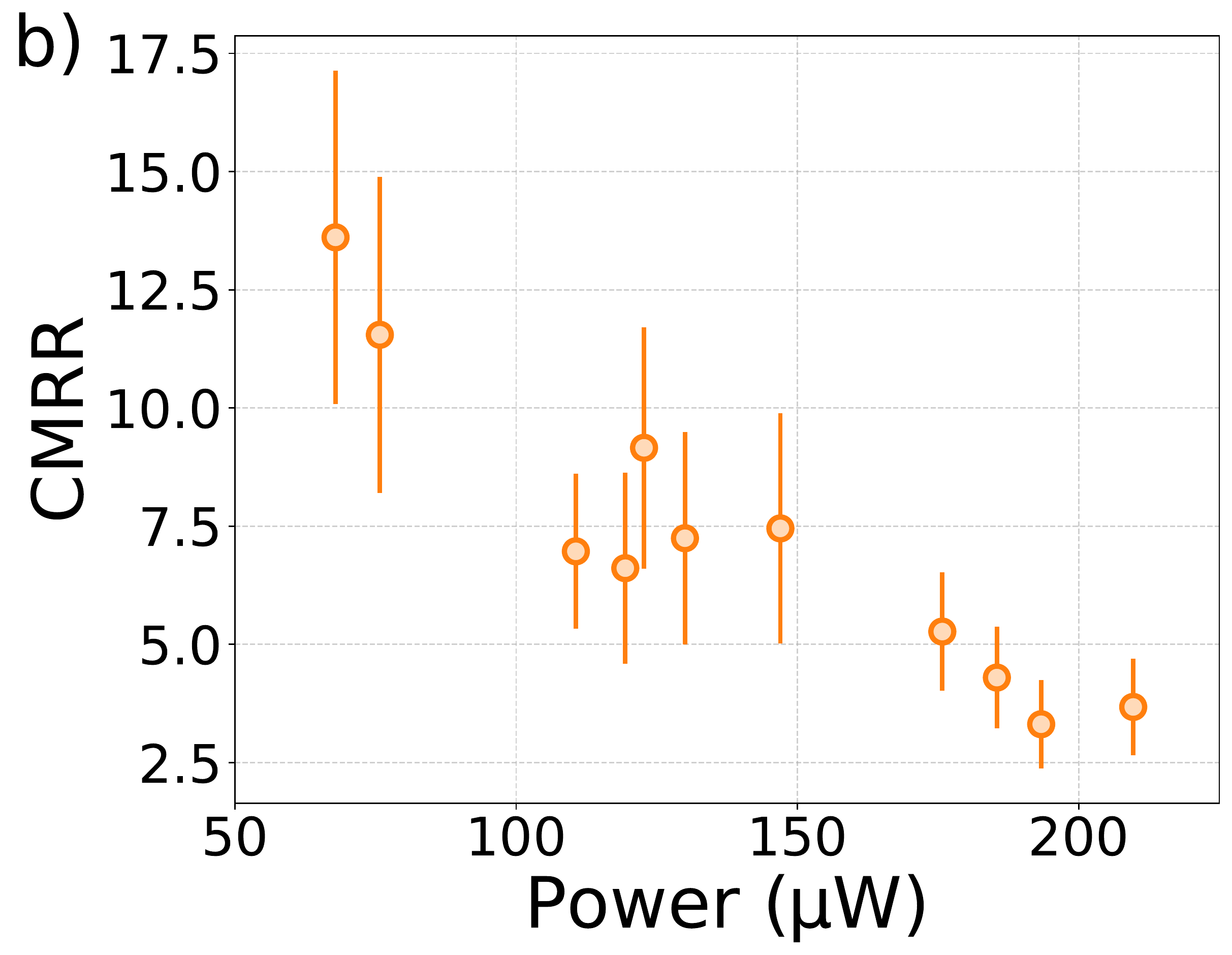}
\caption{CMRR between sum and difference in the measurements of sideband-pump correlation (a) and sideband-sideband correlation (b) vs the power of the single sideband. The laser used is a Fabry-Pérot QCL emitting light at \SI{4.5}{\micro \meter}, working at room-temperature (\SI{23}{\celsius}). The data are obtained spanning the laser current in the range \SIrange[]{512}{517}{mA}. To calculate the CMRR, the sum and difference INPSDs have been averaged in a \SI{1}{MHz}-FFT window centered around \SI{10}{MHz} and the ratio between the two has been computed. The data shown in these graphs refer to the Fabry-Pérot laser emitting at \SI{4.5}{\micro m}.}\label{fig:CMRR}
\end{figure} 
Due to this descending trend, the INPSD is expected to reach the shot-noise level at around a few GHz, suggesting that with a faster detector it would be possible to observe non-classical features of the INPSD itself (sum INPSD below the shot-noise level).  {Fast commercial MIR detectors are currently available (with a bandwidth of a few GHz), but their dynamic range is insufficient to detect the shot noise of the incident power. This is because the detector surface is decreased to get a higher bandwidth resulting in a lower power level of saturation, while the background noise level is the same as larger-area detectors.} To better understand the correlation (anti-correlation) trend measured in our analysis, in \textbf{Figure~\ref{fig:CMRR}} the two types of correlations are represented as functions of the single sideband emission power. In these measurements the temperature of the laser was fixed and the laser current was varied to change the sideband power. We calculated the CMRR as the ratio between the sum INPSD and the difference INPSD obtained with the correlation setup. The graph (b) shows that the CMRR decreases by increasing the power of the two sidebands. We remark that in the case of sideband-sideband correlation, the CMRR is above 1 (CMRR$>1$). 
Moreover, for all the measurements, as suggested by the INPSD shown in Figure~\ref{fig:INPSD_corr_anticorr}, the difference INPSD lies above the corresponding shot-noise level. This means that the noise in the sidebands is partially correlated and that there is still an excess of uncorrelated noise that prevents the achievement of the sub-shot-noise level. In case of sideband-pump correlation we obtained a CMRR~$<1$ suggesting a partial anti-correlation between the two modes (Figure \ref{fig:CMRR} (a)). Even in this case the INPSD remains well above the corresponding shot-noise level. The correlation sign between the modes is schematically shown in \textbf{Table~\ref{tab:correlation_tab}}, where the $+$ sign is associated to correlation, while the $-$ sign to anti-correlation. 
\begin{table}
\begin{center}
\vspace{\baselineskip}
\begin{tabular}{c | c | c | c} 
 & mode +1 & mode -1 & mode 0 \\ \hline 
mode +1 & + & + & $-$ \\ \hline 
mode -1 & + & + & $-$ \\ \hline 
mode 0 & $-$ & $-$ & +  
\end{tabular}
\end{center}
\caption[]{Correlation sign between the modes. The + sign indicates correlation, the $-$ sign indicates anti-correlation. }
\label{tab:correlation_tab}
\end{table}
Interestingly, we can observe from the graphs (Figure~\ref{fig:CMRR}) that the amount of correlation between the two sidebands decreases by increasing the laser gain. In contrast, the anti-correlation between the pump and a sideband increases. These trends suggest that the correlation is higher when the sidebands are close to threshold, while the competition with the central mode gets stronger at increasing power. Therefore, studying the system nearby the sidebands' threshold seems more favorable for unveiling FWM-induced correlation. Indeed, besides the losses, gain and mode competition seem to be the two parameters degrading the correlation between the two sidebands and preventing the detection of any non-classicality. 
\begin{figure}[!htbp]
\centering
\includegraphics[width=0.6\columnwidth]{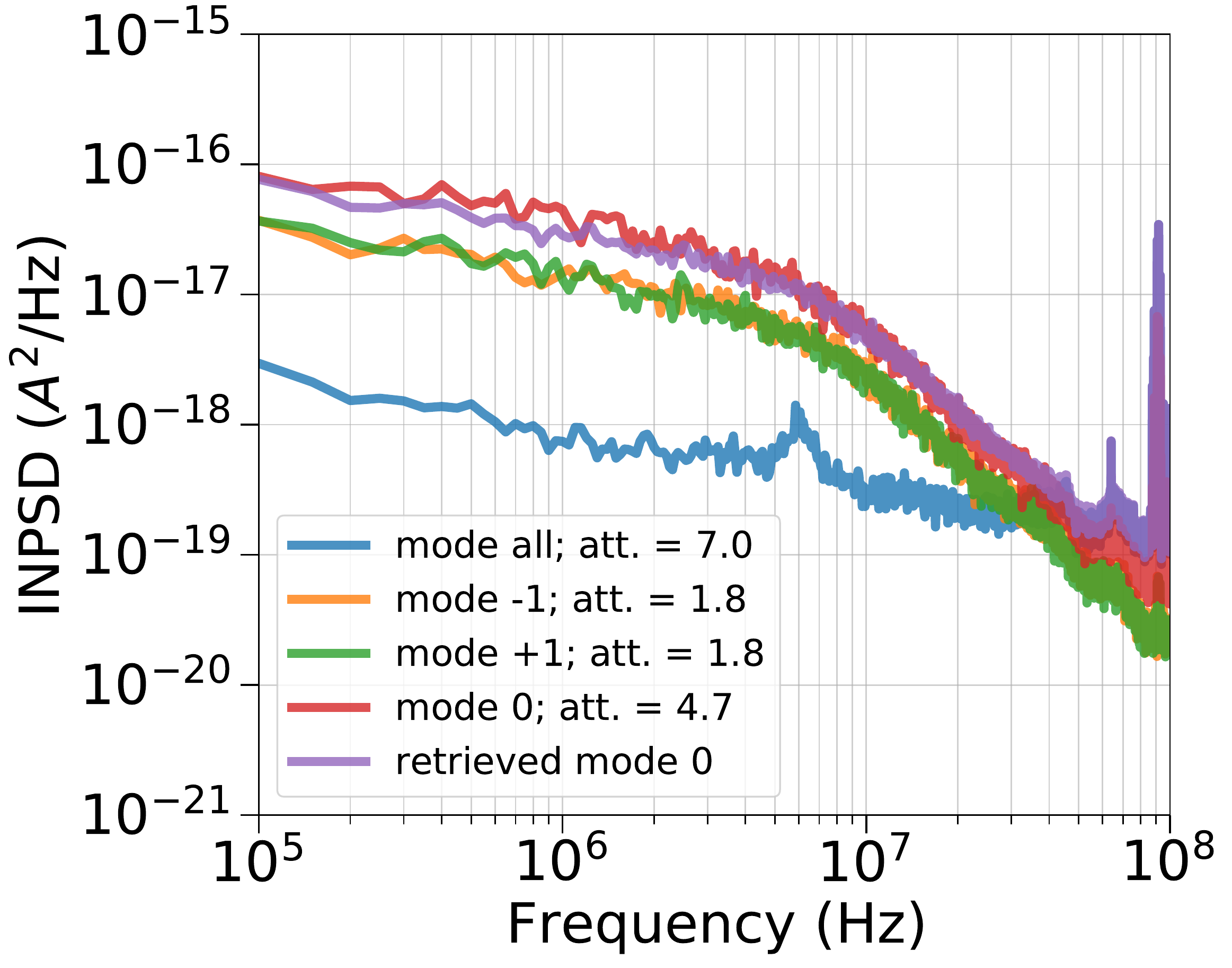}
\caption{INPSD of each individual mode of the three-mode emission and of the whole signal sent to the balanced detector without modal dispersion (no grating). The spectra have been compensated for the related optical attenuation as reported in the legend. The \emph{retrieved mode 0} spectrum has been obtained by adding the spectrum of mode +1, -1 and of the whole signal (\emph{mode all}) according to Equation~\ref{eq:comparison}. The data shown in these graphs refer to the Fabry-Pérot laser emitting at \SI{4.5}{\micro m}. 
}
\label{fig:INPSD_all_comparison}
\end{figure}
All the results shown in this article for the Fabry-Pérot QCL working at \SI{4.5}{\micro \meter} have been observed also for the other lasers, suggesting a common behaviour in the harmonic-comb correlations in QCLs. We remark that even for these correlation (anti-correlation) measurements we were limited from using just two detectors  {for performing more complex investigation with respect to two-mode correlation at a time. T}he performed measurements could highly benefit by using three synchronised detectors, one for each mode. Indeed, the simultaneous acquisition of the three modes allows to retrieve the correlations among the modes in real time, and even to compare the correlations between one mode, e.g. the pump, and the sum of the other two. As  {a valid} alternative, considering  {that} the processes generating the noise  {is} stationary  {and and is reproducible~\cite{note:sim_acq}}, and knowing the correlation signs between the different modes (see Table~\ref{tab:correlation_tab}) we can compare the noise level of the whole signal with the post-processed combination of the single modes, each of them normalized for the applied attenuation (see \textbf{Figure~\ref{fig:INPSD_all_comparison}}). 
The idea is that the noise spectra of all the modes can be added by considering the proper sign (see Table~\ref{tab:correlation_tab}) giving as result the noise of the whole signal: 
\begin{equation}
\text{INPSD}_\text{ 0}-(\text{INPSD}_\text{+1} +\text{INPSD}_\text{-1}) = \text{INPSD}_\text{all} \, .
\label{eq:1}
\end{equation}
For computational reasons we have chosen to invert the equation as follows:  
\begin{equation}
\text{INPSD}_\text{ +1}+\text{INPSD}_\text{-1} +\text{INPSD}_\text{all} = \text{INPSD}_\text{0} \, .
\label{eq:comparison}
\end{equation}
The left-side term of \textbf{Equation~\ref{eq:comparison}} is represented in Figure~\ref{fig:INPSD_all_comparison} (labelled \emph{retrieved mode 0}) and its congruence with the noise spectrum of mode~0 is evident, confirming the consistency of the analysis. \textbf{Equation~\ref{eq:1}} and \textbf{Equation~\ref{eq:comparison}} are valid as long as the sideband-sideband correlation and sideband-pump anti-correlation are dominant with respect to the intrinsic noise of each mode (see~\textbf{Equation~\ref{eq:covariance}}).

\section{Conclusion}

In this work, we have studied the intensity correlations between the different modes that occur in a three-mode harmonic QCL-comb. Our analysis reveals that the two sidebands are correlated and each of them is anti-correlated with the central mode. Furthermore, by increasing the power of the single sideband, the sideband-sideband correlation decreases and, contextually, the pump-sideband anti-correlation increases. This suggests a gain competition between the central mode and the two sidebands that starts to be more and more evident as their power increases. From this first experimental evidence it seems that only minimizing  degradation mechanisms, e.g. losses and gain competition, while enhancing the non-linear interaction, i.e. FWM, will eventually allow to reveal the possible quantum nature of QCLs-comb emission. In the present setup, the INPSD difference in the correlation measurements remains well above the corresponding shot-noise level (over 20~dB).  {Therefore, we were not limited by the ratio between the shot-noise level and the background noise to unveil such correlation in these measurements. In the perspective of observing quantum correlation, the present setup can be successfully used for a single-sideband power above \SI{0.1}{mW}, as suggested by the characterization of the clearance of our detection system reported in reference \cite{Gabbrielli:2021}.} Even though no quantum correlations have been detected, one way of improving the setup in this direction is to work for the development of a new generation of QCLs, where the excess noise is suppressed and, as a consequence, the INPSD of each single mode can be closer to the corresponding shot-noise level. Furthermore, the evidence of a cut-off around \SI{10}{MHz} in the sum INPSD in the correlation measurements, which was confirmed not to depend on the detector, suggests the presence of a dynamics who prevents the INPSD from reaching the shot-noise level. These measurements could highly benefit from using a detection system with a bandwidth of the order of some GHz, which could confirm this hypothesis by allowing detection beyond the cross point between the observed INPSD decreasing trend and the shot-noise level. Reasonably, the bandwidth should not exceed the QCL cavity linewidth, which for a typical 5-mm waveguide gives a frequency limit around \SI{9}{GHz}, and the one of the lasing process \SI{100}{fs} which gives a frequency limit of \SI{10}{GHz}. Another key point in the correlation measurements is the overall quantum detection efficiency, which ideally should be the same for each mode and maximized. Indeed, attenuation hampers the observation of quantum correlations among the simultaneously-detected modes and it severely affects balanced detection when it is uneven. Regarding the detector technology, a further step investigation into quantum effects could be to develop a detection system with ultra-low background noise, that allows to measure the shot-noise level of very weak sidebands with a higher clearance. With the tested state-of-the-art technology, the shot-noise level can be covered by the dark noise in case of very weak sidebands, preventing any quantum signal detection. Finally, we believe that our measurements and the related analysis can substantially contribute to the development of a dedicated model of the intensity correlation among the modes in QCL-combs, which is still missing. In this framework, quantum simulations of electron transport in their active medium could be crucial~\cite{Trombettoni:2021} for identifying which parameters can suitably contribute to a future generation of QCLs with non-classical noise.

In conclusion, our work provides some experimental evidences towards the development of QCL-combs with non-classical emission. Further investigation, both theoretical and experimental, could lead to a revolution in the field of quantum technologies, as it implies the feasibility of direct chip-scale emitters of non-classical states of light, applicable in various fields from quantum sensing to quantum communications.

\section*{Conflict of interest} 
The authors declair no conflict of interest.


\section*{Data Availability Statement} 
The data that support the findings of this study are available from the cor-responding author upon reasonable request.

\section*{Acknowledgements} 
The authors gratefully thank the collaborators within the consortium of the Qombs Project, in particular, the company ppqSense for having provided the ultra-low-noise laser current drivers (QubeCLs). \\ 
The Authors acknowledge financial support from the European Union’s Horizon 2020 Research and Innovation Programme (Qombs Project, FET Flagship on Quantum Technologies grant n. 820419; Laserlab-Europe Project grant n. 871124) and from the Italian ESFRI Roadmap (Extreme Light Infrastructure - ELI Project).


\bibliographystyle{ieeetr}
\bibliography{references}

\begin{thebibliography}{10}
\bibitem{note:sim_acq}
In this type of measurement and comparison, we have to underline an intrinsic
  limitation: in our system, we can do 50/50 single-mode measurements at a time
  or a simultaneous two-mode intensity noise measurement by sending each of the
  two modes to a different detector. In the comparison of the three-mode
  intensity noise we do not have access to all the information given by a
  simultaneous acquisition of all the modes. The simultaneous acquisition of
  the noise of each single sideband and of the central mode would require three
  detectors for the intensity noise measurements and six detectors for the
  shot-noise calibration of each of them. So, the simultaneous acquisition of
  these signals with the whole unresolved three-mode signal would require
  additional detectors and beam splitters. All the acquisitions should also be
  synchronized. The setup will be, therefore, much more complex and expensive.
  We can however state that, given the stability/reproducibility of three-mode
  emission, it is reasonable to compare acquisitions taken at different times.
  This indicates that, albeit with the due limitations, we can still reasonably
  compare the data taken asynchronously.
\end{thebibliography}

\end{document}